\documentclass[superscriptaddress,preprintnumbers,amsmath,amssymb,twocolumn,floats,aps,prl]{revtex4-2}
\usepackage{txfonts}
\usepackage{hyperref}
\usepackage{bm}
\usepackage{xfrac}
\usepackage{amssymb}
\usepackage{graphicx}
\usepackage{amsmath}
\usepackage{epstopdf}
\usepackage{sidecap}
\usepackage{CJK}
\usepackage{url}
\usepackage{color}
\usepackage{amsfonts}
\usepackage{ulem}

\numberwithin{equation}{section}

\begin{document}

\bibliographystyle{revtex4-2}
\title{Photoemission Evidence of a Novel Charge Order in Kagome Metal FeGe}

\author{Zhisheng Zhao}
\thanks{These anthors contributed equally to this work}
\affiliation{School of Emerging Technology and Department of Physics, University of Science and Technology of China, Hefei 230026, China}

\author{Tongrui Li}
\thanks{These anthors contributed equally to this work}
\affiliation{National Synchrotron Radiation Laboratory and School of Nuclear Science and Technology, University of Science and Technology of China, Hefei, 230026, China}

\author{Peng Li}

\affiliation{School of Emerging Technology and Department of Physics, University of Science and Technology of China, Hefei 230026, China}

\author{Xueliang Wu}
\affiliation{Low temperature Physics Laboratory, College of Physics and Center of Quantum Materials and Devices, Chongqing University, Chongqing 401331, China}

\author{Jianghao Yao}
\affiliation{School of Emerging Technology and Department of Physics, University of Science and Technology of China, Hefei 230026, China}

\author{Ziyuan Chen}
\affiliation{School of Emerging Technology and Department of Physics, University of Science and Technology of China, Hefei 230026, China}

\author{Shengtao Cui}
\affiliation{National Synchrotron Radiation Laboratory and School of Nuclear Science and Technology, University of Science and Technology of China, Hefei, 230026, China}

\author{Zhe Sun}
\affiliation{National Synchrotron Radiation Laboratory and School of Nuclear Science and Technology, University of Science and Technology of China, Hefei, 230026, China}

\author{Yichen Yang}
\affiliation{State Key Laboratory of Functional Materials for Informatics, Shanghai Institute of Microsystem and Information Technology (SIMIT), Chinese Academy of Sciences, Shanghai 200050,  China}

\author{Zhicheng Jiang}
\affiliation{State Key Laboratory of Functional Materials for Informatics, Shanghai Institute of Microsystem and Information Technology (SIMIT), Chinese Academy of Sciences, Shanghai 200050,  China}

\author{Zhengtai Liu}
\affiliation{Shanghai Synchrotron Radiation Facility, Shanghai Advanced Research Institute, Chinese Academy of Sciences, Shanghai 201210, China}

\author{Alex Louat}
\affiliation{Diamond Light Source Ltd., Harwell Science and Innovation Campus, Didcot, OX11 0DE, United Kingdom}

\author{Timur Kim}
\affiliation{Diamond Light Source Ltd., Harwell Science and Innovation Campus, Didcot, OX11 0DE, United Kingdom}

\author{Cephise Cacho}
\affiliation{Diamond Light Source Ltd., Harwell Science and Innovation Campus, Didcot, OX11 0DE, United Kingdom}

\author{Aifeng Wang}
\affiliation{Low temperature Physics Laboratory, College of Physics and Center of Quantum Materials and Devices, Chongqing University, Chongqing 401331, China}

\author{Yilin Wang}

\affiliation{School of Emerging Technology and Department of Physics, University of Science and Technology of China, Hefei 230026, China}

\author{Dawei Shen}

\affiliation{National Synchrotron Radiation Laboratory and School of Nuclear Science and Technology, University of Science and Technology of China, Hefei, 230026, China}
\affiliation{New Cornerstone Science Laboratory, University of Science and Technology of China, Hefei, 230026, China}

\author{Juan Jiang}
\email{jjiangcindy@ustc.edu.cn}
\affiliation{School of Emerging Technology and Department of Physics, University of Science and Technology of China, Hefei 230026, China}

\author{Donglai Feng}
\email{dlfeng@ustc.edu.cn}
\affiliation{School of Emerging Technology and Department of Physics, University of Science and Technology of China, Hefei 230026, China}
\affiliation{National Synchrotron Radiation Laboratory and School of Nuclear Science and Technology, University of Science and Technology of China, Hefei, 230026, China}
\affiliation{New Cornerstone Science Laboratory, University of Science and Technology of China, Hefei, 230026, China}
\affiliation{Collaborative Innovation Center of Advanced Microstructures, Nanjing, 210093, China}
\affiliation{Shanghai Research Center for Quantum Sciences, Shanghai, 201315, China}

\date{\today}

\begin{abstract}

A charge order has been discovered to emerge deep into the antiferromagnetic phase of the kagome metal FeGe. To study its origin,  the evolution of the low-lying electronic structure  across the charge order phase transition is investigated with angle-resolved photoemission spectroscopy. We do not find signatures of nesting between Fermi surface sections or van-Hove singularities in zero-frequency joint density of states, and there are no obvious energy gaps at the Fermi level, which exclude the nesting mechanism for the charge order formation in FeGe. However, two obvious changes in the band structure have been detected, i.e., one electron-like band around the $K$ point and another one around the $A$ point move upward in energy position when the charge order forms. These features can be well reproduced by our density-functional theory calculations, where the charge order is primarily driven by magnetic energy saving via large dimerizations of a quarter of Ge1-sites (in the kagome plane) along the $c$-axis.  Our results provide strong  support for this novel charge order formation mechanism in FeGe, in contrast to the conventional nesting mechanism.

\end{abstract}

\maketitle
\clearpage
%---------------------------introduction-------------------------

\begin{SCfigure*}
\includegraphics[width=120 mm]{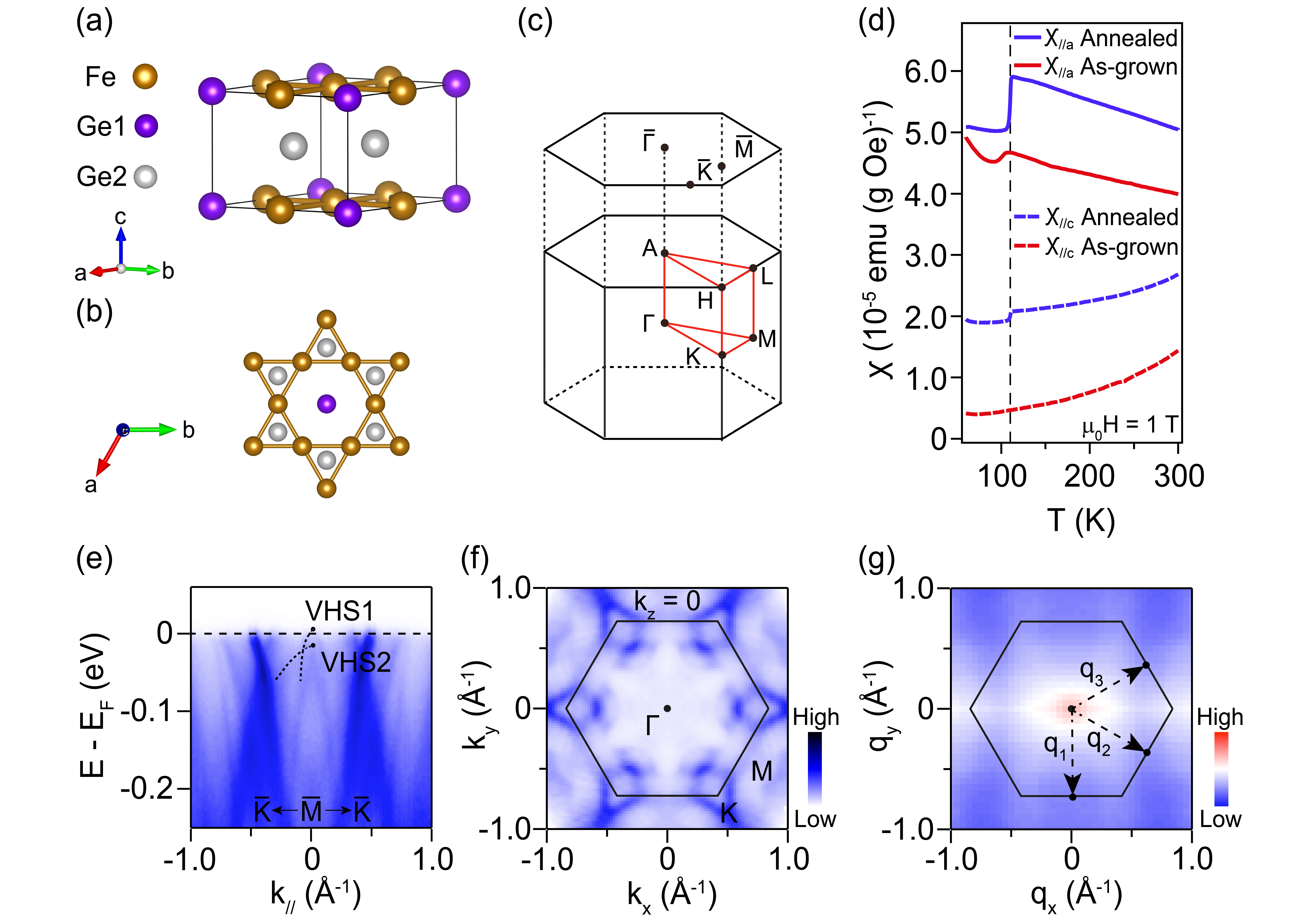}
\caption{ (a) Cystal structure of kagome metal FeGe. Ge1 stands for the position of Ge atom in the kagome layer and Ge2 stands for that in the honeycomb layer. (b) Top view of the crystal structure along the $c$-axis. (c) 3D BZ and its projected 2D surface BZ, with the high-symmetry points indicated by the black dots. (d) Temperature dependent magnetic susceptibility of the as-grown (red curves) and the annealed (blue curves) samples under 1 T magnetic field. Solid lines and dashed lines are measured along the $a$-axis ($\chi$$_/$$_/$$_a$) and $c$-axis ($\chi$$_/$$_/$$_c$), respectively. (e) High symmetry cut along $\overline{\it{K}}$ - $\overline{\it{M}}$ - $\overline{\it{K}}$ acquired at 15 K with 95 eV photons. Black dashed lines indicated the positions of VHS1 and VHS2. (f) Symmetrized Fermi surface map acquired at 130 K with 65 eV photons, which is around the $k_z$ = 0 plane. (g) The zero-frequency JDOS obtained based on (f). }
\label{Fig 1}
\end{SCfigure*}

Compounds with kagome lattice offer a fertile playground to investigate novel electronic states, such as quantum spin liquid~\cite{balents2010spin, banerjee2018excitations, norman2016colloquium},  fractional Chern insulator state~\cite{tang2011high, zhu2016interaction}, charge order (CO)~\cite{guo2009topological, ortiz2020cs, arachchige2022charge} and superconductivity~\cite{ko2009doped, wang2013competing}, due to the  geometric frustration~\cite{syozi1951statistics, song2019unifying, norman2016colloquium}, flat band~\cite{tang2011high, kang2020dirac, okamoto2022topological}, and non-trivial band topology~\cite{ortiz2020cs, ye2018massive, zhou2022chern} in this unique structure. Amongst them, various exotic COs are of particular interest, since they are often intertwined with  other forms of order, including nematicity~\cite{nie2022charge}, superconductivity~\cite{yu2021unusual, chen2021double, luo2023unique} and chirality~\cite{jiang2021unconventional}. One example is the family of AV$_3$Sb$_5$ (A = K, Rb, and Cs)~\cite{ortiz2019new} with vanadium kagome nets, in which chiral charge density wave (CDW)~\cite{jiang2021unconventional, shumiya2021intrinsic, feng2021chiral}, unidirectional charge stripe order~\cite{li2023unidirectional, zhao2021cascade, xu2021multiband} and superconductivity with a pair density wave order~\cite{chen2021roton} have been unveiled.

Recently, another kagome metal FeGe has drawn considerable attention owing to its intimate correlation between the $2\times 2\times 2$ CO phase ($T_{CO}\sim100$ K) and the A-type antiferromagnetic (AFM) phase ($T_{N}\sim~410$ K) therein~\cite{setty2022electron, yin2022discovery, teng2022discovery, shao2023intertwining, miao2022charge, teng2023magnetism, zhou2022magnetic, wu2023novel, ma2023theory, chen2023charge, wang2023enhanced, duan2022epitaxial, chen2023longranged, wu2023annealing}. The recent angle-resolved photoemission spectroscopy (ARPES) study have found energy gaps near the Fermi level ($E_F$), and suggested that the nesting among the van-Hove singularities (VHSs) near $E_F$ drives CO~\cite{teng2023magnetism}.
Moreover, scanning tunneling microscopy (STM) studies observe a short-ranged $2\times 2\times 2$ CO in as-grown FeGe~\cite{yin2022discovery, teng2022discovery, chen2023charge}, and a long-ranged CO in annealed sample with much less defects on Ge2-sites (in the honeycomb plane) ~\cite{chen2023longranged, wu2023annealing}. In the as-grown samples, X-ray diffraction (XRD) reveals the first order structural transition associated with CO~\cite{miao2022charge}, while the refinement of XRD data become possible for high-quality annealed FeGe samples, which find that one quarter of the Ge atoms on the Ge1-site (in the kagome plane) are strongly dimerized along $c$-axis at the structural transition~\cite{chen2023longranged}. Theoretical calculations have predicted such a dimerization, and in fact, it has been shown that the magnetic energy is saved in this process, which is the primary driving force for the structural transition and CO~\cite{miao2022charge, wang2023enhanced}, in contrast to the electronic energy saving process in a conventional nesting mechanism for CDW. Consistently, negative phonon frequencies are not found in the non-CO state ~\cite{miao2022charge}, distinct from the case in AV$_3$Sb$_5$~\cite{tan2021charge, ratcliff2021coherent, christensen2021theory}.
The contradiction between these findings and the early ARPES data
thus calls for an investigation of the electronic structure of the annealed FeGe samples with long-ranged CO, which could help to clarify the CO mechanism in this kagome magnet.

In this Letter, we study the evolution of the low-lying electronic structure of the annealed FeGe single crystals  across the CO phase transition with ARPES  and DFT calculations. Distinct from previous ARPES results, we do not find noticeable energy gap on the  Fermi surface, suggesting that the nesting mechanism of conventional CDW is not  responsible for the charge ordering in FeGe. Instead, we discover that the spectral shift upward in energy position around both the $K$ and $A$ points in the Brillouin zone (BZ) as the sample enters the CO state, which costs electronic energy. Our findings can be well reproduced by the DFT calculations that find the CO phase transition in FeGe is driven by lowering magnetic exchange energies via the large dimerization ($\sim$ 1.3 \AA) of a quarter of Ge1-sites along the $c$-axis~\cite{miao2022charge, wang2023enhanced}. Our ARPES data thus support this novel CO mechanism based on magnetic-energy saving.
\begin{figure*}
\includegraphics[width=180 mm]{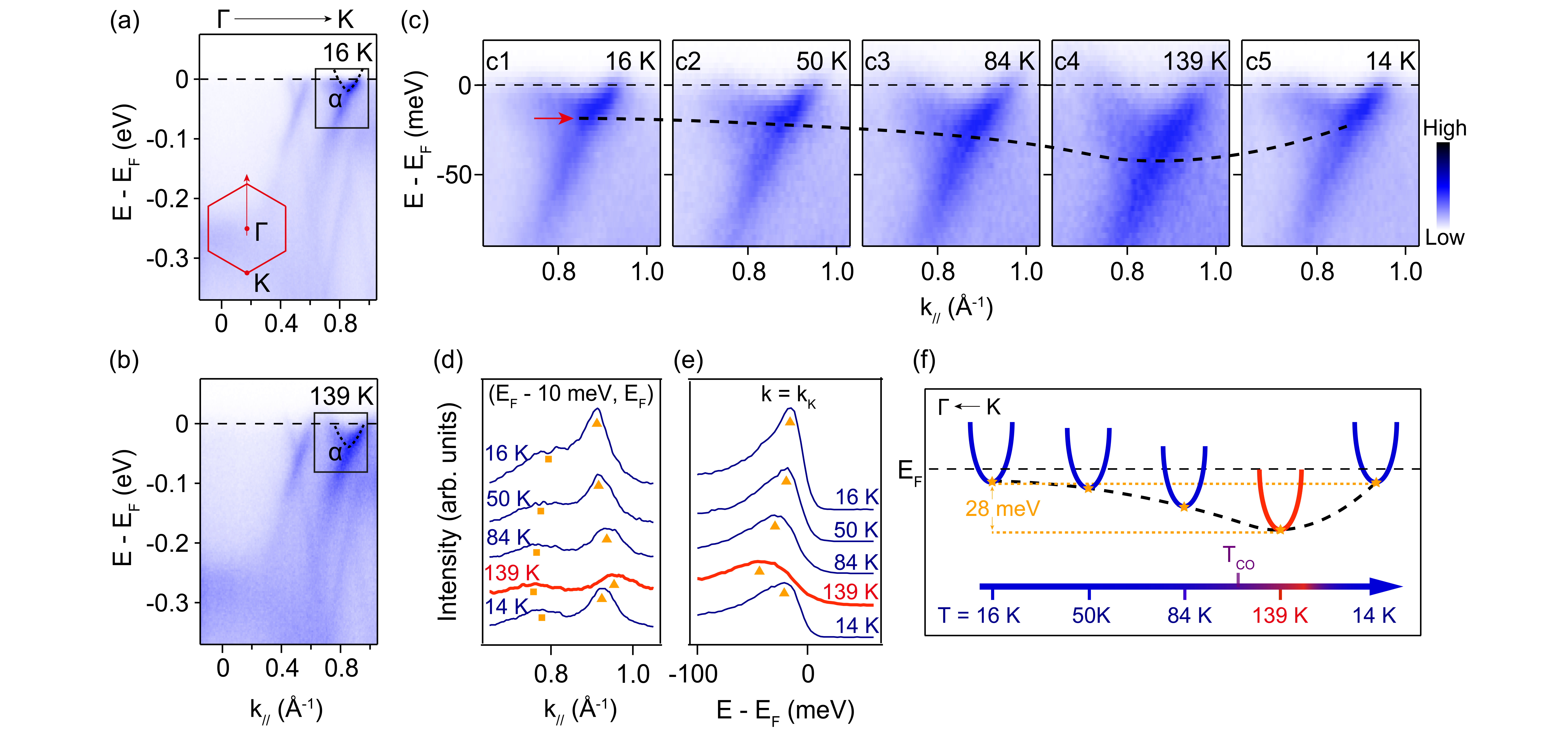}
\caption{ (a) Electronic structure along  $\mit{\Gamma}$ - $K$ measured at 16 K. (b) Similar to (a) but measured at 139 K. Black dashed line indicates the $\alpha$ band position. (c) Zoom in spectra of the $\alpha$ band in the region highlighted by the black frame in (a) and (b) measured at different temperatures crossing $T_{CO}$. (d) Corresponding integrated MDCs acquired at ($E_F$ - 10 meV, $E_F$). The yellow squares and triangles indicate the momentum positions of Fermi crossing at each temperatures. (e) Corresponding integrated EDCs acquired at ($k_K$ - 0.015 \AA$^-$$^1$, $k_K$ + 0.015 \AA$^-$$^1$), where $k_K$ represents the momentum position of the $K$ point. The yellow triangles indicate the energy positions of the band bottom of $\alpha$ at each temperatures. (f) A schematic illustrating the temperature-dependent evolution of the $\alpha$ band. All the data are measured with 23 eV photons.}
\label{Fig 2}
\end{figure*}
\begin{figure*} [htbp]
\includegraphics[width=18 cm]{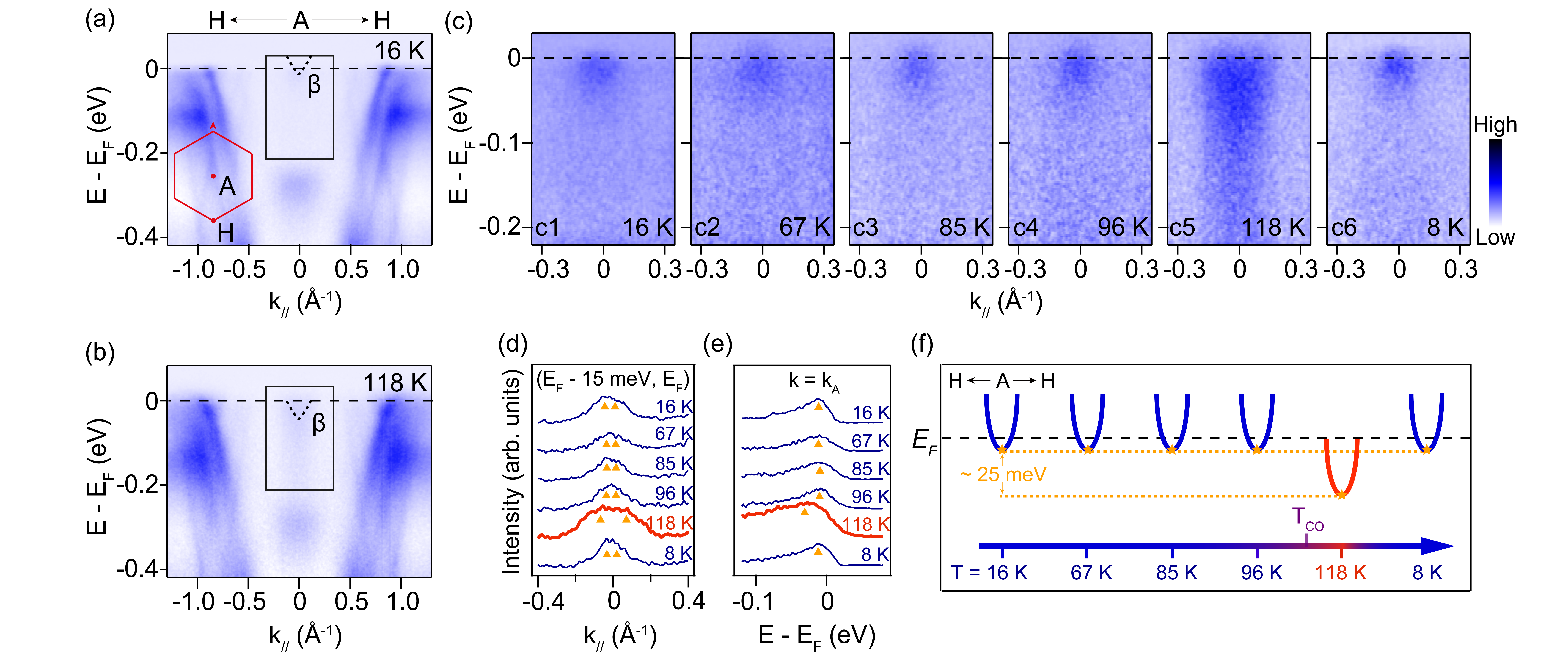}
\caption{ (a) Electronic structure along $H$ - $A$ - $H$ measured at 16 K.  (b)  Similar to (a) but meaured at 118 K. Black dashed line indicates the $\beta$ band position. (c)  Zoom in spectra of the $\beta$ band in the region highlighted by the black frame in (a) and (b) measured at different temperatures crossing $T_{CO}$. (d) Corresponding integrated MDCs acquired at ($E_F$ - 15 meV, $E_F$).  The yellow squares and triangles indicate the momentum positions of Fermi crossing at each temperatures. (e) Corresponding integrated EDCs acquired at ($k_A$ - 0.1 \AA$^-$$^1$, $k_A$ + 0.1 \AA$^-$$^1$), where $k_A$ represents the momentum position of the $A$ point. The yellow triangles indicate the energy positions of the band bottom of $\beta$ at each temperatures. (f) A schematic illustrating the temperature-dependent evolution of the $\beta$ band. All the data are measured with 95 eV photons.}
\label{Fig 3}
\end{figure*}

Kagome metal FeGe crystallizes in the space group $P6/mmm$ with lattice constants $a$ = 4.99 \AA~and $c$ = 4.05 \AA~\cite{bernhard1988magnetic, meier2020flat}. It consists of Fe$_3$Ge kagome layers and Ge2 honeycomb layers stacking alternatively along the $c$-axis as shown in Figs. 1(a) and 1(b)~\cite{setty2022electron, shao2023intertwining, yin2022discovery, teng2022discovery}. Its corresponding three-dimensional (3D) BZ and projected (001) surface BZ are illustrated in Fig. 1(c), where the black dots represent the high-symmetry momenta. Previous results of STM, neutron scattering and magnetic susceptibility indicate a short-ranged CO in the as-grown FeGe samples~\cite{chen2023charge, teng2022discovery}, which is also confirmed by our temperature-dependent magnetic susceptibility result. It only shows an anomaly in $\chi$$_{\mathop{//}a}$ at the CO onset temperature ($T_{CO}$) [Fig. 1(d)]. Notably, the annealed sample shows sharper transitions in both $\chi$$_{\mathop{//}a}$ and $\chi$$_{\mathop{//}c}$ at $T_{CO}$, which is consistent with a first order transition behavior, indicative of a long-ranged CO therein.

The high-quality single crystals of FeGe were grown using the chemical vapor transport (CVT) method. The as-grown single crystals were further annealed to obtain long-ranged CO \cite{chen2023longranged, wu2023annealing}. ARPES experiments were conducted at beamline 13U of National Synchrotron Radiation Laboratory (NSRL) equipped with a Scienta DA30 electron analyzer, beamline 03U of Shanghai Synchrotron Radiation Facility (SSRF) equipped with a Scienta DA30 electron analyzer and the high resolution branch of beamline I05 of Diamond Light Source (DLS) with a Scienta-Omicron R4000 electron analyzer. The overall energy resolution for the combined instruments was set better than 15 meV, and the angluar resolution was better than 0.1 degrees. The samples were cleaved in situ and measured in an ultrahigh vacuum better than 1.0 $\times$ $10^{-10}$ mbar. Figure 1(e) shows the photoemission intensity along $\overline{K}$ - $\overline{M}$ - $\overline{K}$ taken from the annealed sample, where there are two VHSs assigned as VHS1 (slightly above $E_F$) and VHS2 ($\sim$~15 meV below $E_F$) existing around the $\overline{M}$ point, consistent with previous reports~\cite{teng2022discovery, teng2023magnetism}. However, our detailed temperature-dependent ARPES measurements reveal negligible sign of gap opening at $E_F$, nor obvious VHS shifting as reported in other work~\cite{teng2023magnetism} (see Supplementary Figs. S1 and S2 for more details). It should be noted that we have performed a comprehensive real-space mapping of the ARPES intensity and probed seven points on the cleaved sample [Supplementary Figs. S2], which should effectively cover the CO domains~\cite{chen2023charge, chen2023longranged}. Besides, in Fig. 1(g), we present the zero-frequency joint density of states (JDOS) ~\cite{cho2021emergence, shen2007novel, chatterjee2006nondispersive, mcelroy2006elastic, shen2008primary} based on the symmetrized Fermi surface map detected above the CO transition temperature as displayed in Fig.~1(f). It does not show any pronounced peaks related to the CO wave-vectors $\boldsymbol{q}_{i}$  ($i$ = 1, 2 and 3). This is also confirmed by our calculated zero-frequency JDOS [Supplementary Fig. S3] and consistent with the previous theoretical work~\cite{wu2023novel}. This finding, together with the absence of obvious gap opening around the $\overline{M}$ point, indicate that the nesting mechanism might not be the dominating driving force for the CO formation of FeGe.

To identify the states closely associated with the CO formation, we conducte a comprehensive search for the spectral changes across the CO phase transition over the whole 3D BZ. According to the detailed photon-energy-dependent ARPES measurements, one can identify that the data taken with  23 eV and 69 eV photons go through the $k_z$ = 0 plane while the data taken with 44 eV and 99 eV photons go through the $k_z$ = $\pi$ plane, with an inner potential of 17 eV [Supplementary Fig. S4]. After an extensive mapping of the electronic structure over the BZ, two obvious spectral changes accompanied with the long-ranged CO formation are resolved: one is around the $K$ point and the other is around the $A$ point in the BZ.

Figure 2(a) shows the photoemission intensity plot along $\mit\Gamma$ - $K$ acquired at 16 K, in which one electron pocket around $K$ (assigned as $\alpha$), with a band bottom at $\sim$ 15 meV below $E_F$ can be observed. Compared to the non-CO state at $T=139$ K, $\alpha$ band moves upward by $\sim$ 28 meV in energy position in the CO state [Fig. 2(b)]. The detailed enlarged spectral evolution in the region around $K$ [marked by the black frame in Fig.~2(a)] with temperature crossing $T_{CO}$ is shown in Fig.~2(c). A constant sinking of the $\alpha$ band in energy position with increased temperature can be observed. And the $\alpha$ band is temperature-independent in the non-CO state, as demonstrated in Supplementary Fig. S5. Notably, the surface aging effect is negligible, since the last data taken at 14 K is similar to that of 16 K taken in the beginning. Moreover, the integrated momentum distribution curves (MDCs) around $E_F$ [Fig.~2(d)], together with the integrated energy distribution curves (EDCs) around $K$ [Fig.~2(e)] further highlight such a continuous spectral change. This behavior is summarized schematically in Fig.~2(f), which can be deduced from the dispersion extracted based on the second-derivative of the photoemission intensity, and the Fermi surface maps around $K$ as well [Supplementary Fig. S6].

\begin{figure}[htbp]
\includegraphics[width=87 mm]{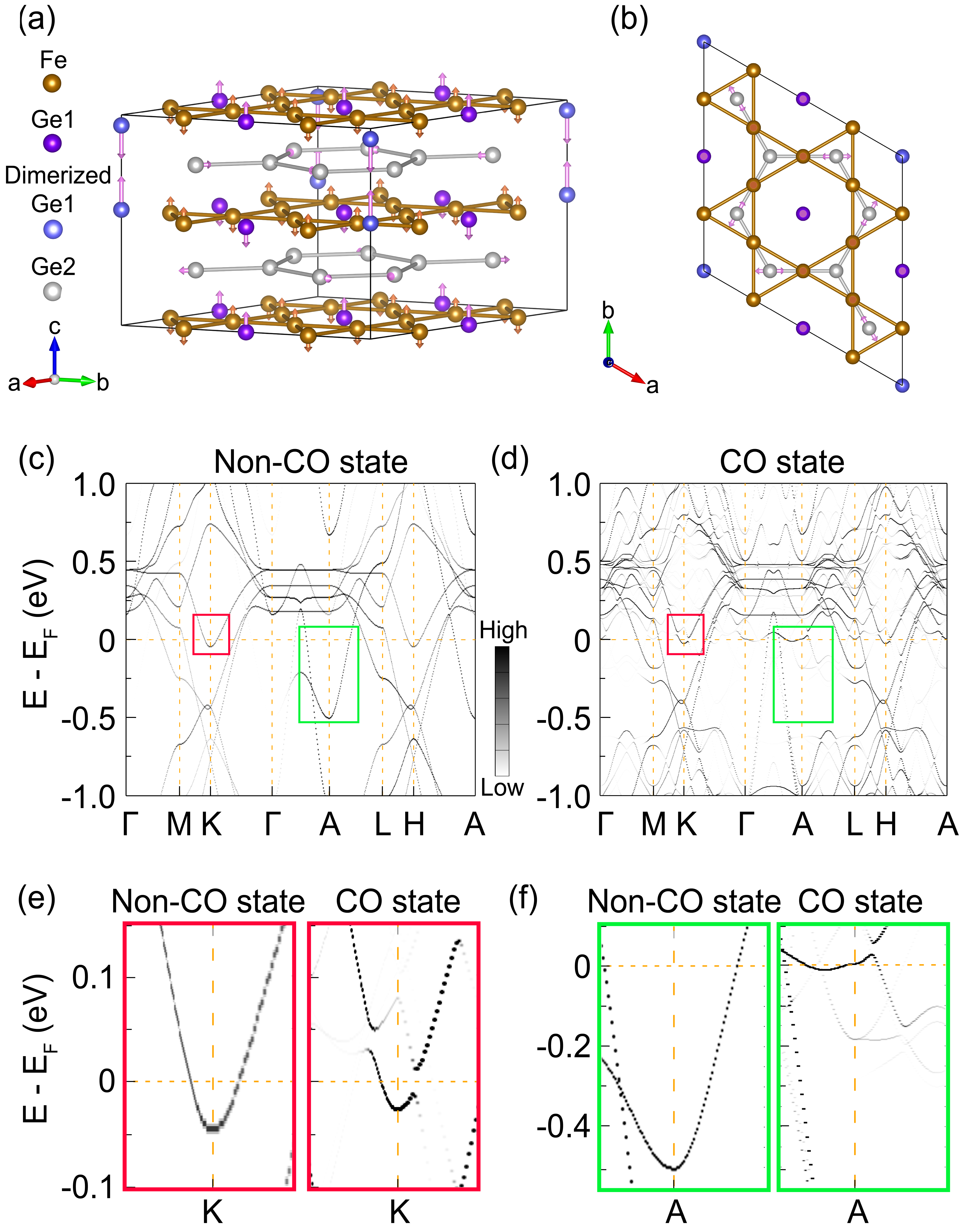}
\caption{ (a) The 2 $\times$ 2 $\times$ 2 CO superstructure of FeGe, including a large dimerization of a quarter of Ge1-sites along the $c$-axis (indicated by the longer pink arrows) and small distortions of Fe-sites along the $c$-axis (yellow arrows) as well as a Kekul\'{e}-type distortions of Ge2-sites (shorter in-plane pink arrows). (b) Top view of the dimerized structure along the $c$-axis. (c), (d) The DFT-calculated band structures in the non-CO and CO states, respectively. It should be noted that the band structures have been unfolded into the 1 $\times$ 1 $\times$ 1 non-magnetic BZ. (e), (f) Highlights of the Fe-$3d$ ($\alpha$) and Ge-$4p$ ($\beta$) bands around the $K$ and $A$ points, respectively.}
\label{Fig 4}
\end{figure}

A more dramatic spectral change across the CO phase transition is discovered in the $k_z$ = $\pi$ plane. As shown in Fig.~3(a), there exists a tiny electron-like band (assigned as $\beta$) centered at the $A$ point, which barely crosses $E_F$ in the CO state at $T=16$ K. In contrast, the bottom of $\beta$ shifts downward in energy position ($\sim$~25 meV) in the non-CO state at $T=118$ K [Fig.~3(b)]. The detailed temperature-dependent evolution of the $\beta$ band is shown in Fig.~3(c). It should be noted that, since the intensity of the $\beta$ band is rather weak, all spectra here have been subtracted by a smoothed background EDC to highlight their characteristics (see more details in Supplementary Fig. S7). One can clearly identify a sudden change of the $\beta$ band when crossing $T_{CO}$, which is further proved by the corresponding MDCs at $E_F$ [Fig.~3(d)], EDCs around A [Fig.~3(e)], and second-derivative data and Fermi surface maps around $A$ [Supplementary Fig.~S7], respectively. Besides, the temperature-independent behavior of the $\beta$ band in the non-CO state is presented in Supplementary Fig. S8. Accordingly, we summarize a schematic of temperature-dependent evolution of $\beta$ in Fig.~3(f).

To understand the origin of the observed band structure evolution across $T_{CO}$, we turn to our recent DFT calculations ~\cite{miao2022charge, wang2023enhanced}, which predicte a large dimerization of a quarter of Ge1-sites along the $c$-axis in a $2\times 2\times 2$ superstructure. Figures 4(a) and 4(b) show the $2\times 2\times 2$ CO superstructure relaxed by DFT. The large partial Ge1-dimerization induces CO in the kagome and honeycomb layers accompanied with small distortions of Fe-sites along the $c$-axis and a Kekul\'{e}-type distortion of Ge2-sites. Here, we calculate and  compare the band structures between the non-CO and CO states to extract the low-lying electronic structure changes across the phase transition, which are shown in Figs. 4(c) and 4(d), respectively. The band structures have been unfolded into a $1\times 1\times 1$ non-magnetic BZ. Obvious differences can be found around the $K$ and $A$ points near $E_F$. Figure 4(e) highlights the band structure around the $K$ point, where the electron pocket with Fe-$3d$ character shifts upward in the CO state, accompanied by a small hybridization gap opening due to the band folding effect. This change is attributed to the small distortions of Fe-sites. The more dramatic change is at the $A$ point, where the electron pocket with Ge-$4p$ character shifts upward a lot in the CO state, as highlighted in Fig. 4(f). This change is attributed to the large Ge1-dimerization.

These two changes in band structure are qualitatively consistent with our ARPES observations despite that the small hybridization gap at the $K$ point is beyond our experimental resolution. The sudden and dramatic change of the Ge-$4p$ electron pocket around the $A$ point observed by ARPES is thus a manifestation of the first-order structural phase transition dominated by the large Ge1-dimerization. Since the CO-induced distortions of Fe-sites are very small and the change of Fe-$3d$ electron pocket around the $K$ point is tiny, it is hard to identify a sudden first-order-like change in this band. One may wonder these two changes both raise up the electronic energy of the system, which seems to contradict with the formation of a CO. However, according to the theory~\cite{wang2023enhanced}, the Ge1-dimerization significantly enhances the spin polarization of the Fe-sites and lowers the magnetic energy, which overcomes the energy cost from structural distortion and leads to the CO ground state. Since there is neither obvious CO gap opening around $E_F$ nor zero-frequency JDOS evidencing nesting behavior, the pure charge degree of freedom could not be responsible for such a 100 K CO phase transition. Taking together, our results support this CO mechanism suggested by the DFT calculations for FeGe.

In conclusion, our ARPES experiments observe neither signatures of nesting of Fermi surfaces or van-Hove singularities nor sizeable electronic energy gaps around the Fermi level that can induce a 100 K CO, which excludes a conventional CDW mechanism driven by saving electronic energies for the CO in FeGe.  Instead, we find the sudden upward shift of an electron-like band dominated by Ge-$4p$ orbitals around the $A$ point, as well as the upward shift of another electron-like band dominated by Fe-$3d$ orbitals around the $K$ point after the CO phase transition in the ARPES spectra, which are well reproduced by our DFT calculation based on the theoretical model in Ref.~\cite{wang2023enhanced}. Thus our ARPES data are consistent with the unconventional CO mechanism proposed in~\cite{wang2023enhanced}, where the CO in FeGe is driven by primarily saving magnetic energies via a first-order structural transition involving large partial Ge1-dimerization along the $c$-axis. Our findings thus clarify the current controversy and make essential steps toward understanding the intertwined magnetic and charge orders in the kagome metal FeGe.

We acknowledge the support by the National Natural Science Foundation of China (Grant No. 12174362, No. 11888101, No. 11790312, No. 92065202 and No. 12174365), the Innovation Program for Quantum Science and Technology (Grant No. 2021ZD0302800), the New Cornerstone Science Foundation, the Innovation Program for Quantum Science and Technology (Grant No. 2021ZD0302800) and the Fundamental Research Funds for the Central Universities of China (WK9990000103). Part of this research used Beamline 03U of the Shanghai Synchrotron Radiation Facility, which is supported by ME2 project under contract No. 11227902 from National Natural Science Foundation of China.

\bibliography{FG}

\bibliographystyle{apsrev4-2}

\end{document}